\documentclass[sigplan,10pt]{acmart}
\pdfoutput=1

\setcopyright{none}

\usepackage{booktabs}   
\usepackage{subcaption} 
\usepackage[ruled,vlined,linesnumbered]{algorithm2e}
\usepackage{multirow}
\usepackage{ragged2e}
\usepackage{caption}

\begin{document}

\title[]{DiT-HC: Enabling Efficient Training of Visual Generation Model DiT on HPC-oriented CPU Cluster}        

\author{Jinxiao Zhang}
\affiliation{
  \department{Department of Earth System Science}              
  \institution{Tsinghua University}            
  \city{Beijing}
  \country{China}                    
}
\email{zhang-jx22@mails.tsinghua.edu.cn}          

\author{Yunpu Xu}
\affiliation{
  \department{Institute of Data and Information}              
  \institution{Tsinghua Shenzhen International Graduate School}            
  \city{Shenzhen}
  \country{China}                    
}
\email{xuyp25@mails.tsinghua.edu.cn}          

\author{Xiyong Wu}
\affiliation{
  \department{Institute of Data and Information}              
  \institution{Tsinghua Shenzhen International Graduate School}            
  \city{Shenzhen}
  \country{China}                    
}
\email{wuxiyong25@mails.tsinghua.edu.cn}          

\author{Runmin Dong}
\authornote{Corresponding Authors}
\affiliation{
  \department{School of Artificial Intelligence}              
  \institution{Sun Yat-sen University}            
  \city{Zhuhai}
  \country{China}                    
}
\email{dongrm3@mail.sysu.edu.cn}          

\author{Shenggan Cheng}
\affiliation{
  \institution{National University of Singapore}            
  \city{Singapore}
  \country{Singapore}                    
}
\email{shenggan@comp.nus.edu.sg}          

\author{Yi Zhao}
\affiliation{
  \department{Department of Earth System Science}              
  \institution{Tsinghua University}            
  \city{Beijing}
  \country{China}                    
}
\email{zhang-y19@mails.tsinghua.edu.cn}          

\author{Mengxuan Chen}
\affiliation{
  \department{Department of Earth System Science}              
  \institution{Tsinghua University}            
  \city{Beijing}
  \country{China}                    
}
\email{chenmx21@mails.tsinghua.edu.cn}          

\author{Qinrui Zheng}
\affiliation{
  \institution{National Supercomputing Center in Shenzhen}            
  \city{Shenzhen}
  \country{China}                    
}
\email{zhengqr@nsccsz.cn}          

\author{Jianting Liu}
\affiliation{
  \institution{National Supercomputing Center in Shenzhen}            
  \city{Shenzhen}
  \country{China}                    
}
\email{liujt@nsccsz.cn}          

\author{Haohuan Fu}
\authornotemark[1]
\affiliation{
  \department{Institute of Data and Information}              
  \institution{Tsinghua Shenzhen International Graduate School}            
  \city{Shenzhen}
  \country{China}                    
}
\email{haohuan@tsinghua.edu.cn}          

\fancyhead{}  
\renewcommand\footnotetextcopyrightpermission[1]{} 

\begin{abstract}

Generative foundation models have become an important tool for data reconstruction and simulation in scientific computing, showing a tight integration with traditional numerical simulations. At the same time, with the development of new hardware features, such as matrix acceleration units and high-bandwidth memory, CPU-based clusters offer promising opportunities to accelerate and scale such models, facilitating the unification of artificial intelligence and scientific computing.
We present DiT-HC, the first system to train and scale the generative model DiT on a next-generation HPC CPU cluster. DiT-HC introduces three key techniques: (1) communication-free tensor parallelism (CFTP) with AutoMem for automated memory-aware dataflow, (2) HCOps, a suite of optimized GEMM and operator kernels leveraging vector and matrix acceleration units, and (3) a custom MPI backend that overlaps computation, communication, and memory movement. Experiments show 8.2$\times$–87.7$\times$ speedups over native or public CPU libraries and 90.6\% weak scaling efficiency on 256 nodes. These results demonstrate the feasibility of large-scale generative model training on CPU clusters and provide new insights for future HPC-AI co-design.

\end{abstract}

\maketitle

\section{Introduction}

Artificial intelligence is becoming increasingly prevalent in scientific research, and scientific computing is shifting from a paradigm centered solely on numerical simulation to one where simulations and data-driven models co-develop in a complementary manner\cite{silvano2025survey,ejarque2022enabling,jiang2024efficient}. While this shift enhances modeling capability\cite{bracco2025machine} and improves the efficiency of data utilization, it also places new demands on high-performance computing (HPC) platforms: they must not only continue to support large-scale simulations in domains such as climate and seismology, but also run generative foundation models efficiently to enable data-driven optimization in areas such as data assimilation\cite{huang2024diffda,xu2025fuxi}, parameterization schemes\cite{chen2023resu,duan2025ai}, and bias correction\cite{kim2021deep,gregory2024machine}. Consequently, HPC platforms are supposed to support both numerical simulation and model training to achieve a unified workflow.

HPC-oriented CPU cluster is one of the most established and widely deployed computing infrastructures in modern supercomputers, forming the backbone of large-scale numerical simulations and scientific applications. In generative model training, GPUs/TPUs are the primary platforms\cite{li2022ai}, while CPU clusters provide only limited support for efficient training of deep learning models. The primary reason for this gap lies in the general-purpose nature of CPUs: their architectures emphasize versatility but typically provide lower compute density and memory bandwidth than GPUs. For deep learning workloads dominated by large-scale matrix operations and intensive memory access workloads, these limitations directly constrain training efficiency, placing CPUs at a disadvantage in scaling generative models and posing challenges unifying AI and HPC on the same platform.

Consequently, there is a trend for next-generation HPC-oriented CPU architectures to incorporate AI-focused hardware features, such as matrix acceleration units (MAUs) and on-package high-bandwidth memory (OPM), which provide new opportunities for deep learning model training. For example, the Armv9 architecture introduces two-dimensional ZA registers and the Scalable Matrix Extension (SME) instruction set, significantly improving the efficiency of matrix and tensor computations \cite{weidmann2021introducing}. Intel’s Sapphire Rapids processors similarly integrate Advanced Matrix Extensions (AMX), enabling more efficient execution of AI workloads \cite{kim2024exploiting}. In addition, modern HPC CPUs are adopting hierarchical memory structures that combine on-package high-bandwidth memory with traditional DDR, alleviating the bandwidth bottlenecks of conventional designs. However, fully exploiting these new features is non-trivial: unlike the highly encapsulated GPU ecosystem with its mature toolchains, CPU platforms involve distinct characteristics in instruction set extensions, hierarchical memory management, and NUMA-domain interconnects, which necessitate systematic rethinking of operator implementations, parallelization strategies, and memory scheduling.

In this context, we select the generative foundation model DiT as our case study to investigate efficient training and scalable optimization on HPC-oriented CPU clusters. On the one hand, diffusion models have become the dominant paradigm in visual content generation \cite{sohl2015deep,ho2020denoising,song2020score,nichol2021improved,lu2022dpm,podell2023sdxl,rombach2022text}, and Transformer-based DiT \cite{Peebles2022DiT,esser2024scaling} serves as a representative with strong global modeling capacity and scalability. On the other hand, diffusion models have also demonstrated significant value in scientific domains \cite{kazerouni2023diffusion}, achieving advances in satellite image super-resolution \cite{dong2024building}, multimodal image fusion \cite{bao2023one}, ensemble weather forecasting \cite{price2023gencast}, and Earth system data assimilation \cite{hess2025fast,huang2024diffda,springenberg2025diffscale}. Therefore, DiT combines the representativeness of cutting-edge generative models with practical relevance in scientific applications, making it a representative target for evaluating and optimizing next-generation CPUs in large-scale training.

In this paper, we introduce DiT-HC, a scalable high performance training framework for DiT on a new HPC-oriented CPU cluster, the LS pilot system with LX2 CPUs. 
DiT-HC is elaborately designed to exploit the hierarchical memory and matrix acceleration hardware of LX2, enabling scalable and high-throughput training. We validate DiT-HC using remote sensing datasets and show its cross-domain transferability, and high parallel efficiency. Experimental results show DiT-HC achieves a 8.2$\times$ performance improvement over the native high-performance software stack and achieves 90.6\% weak scaling efficiency when scaling to 256 nodes. 

Our main contributions are summarized as follows:

\begin{itemize}

    \item[1.] We optimized DiT training on a HPC-oriented CPU cluster, improving efficiency and scalability, and validated its correctness and transferability on remote sensing data.

    \item[2.] We propose a communication-free tensor parallelism method named CFTP and present an automatic memory dataflow management module AutoMem to fully utilize the hierarchical OPM-DDR memory structure.

    \item[3.] We release an extended PyTorch operator math library named HCOps, which includes optimized matrix tiling schemes for GEMM and low-level intrinsic optimizations for AI operators to better exploit the vector and matrix acceleration units.

    \item[4.]  To reduce resource contention, we implement a custom communication backend supporting MPI asynchronous collective communication and overlap computation, communication, and memory operations across different CPU cores.

\end{itemize}

\section{Related Works}

Several attempts have explored deep learning training on CPU-based supercomputers. The Fugaku system \cite{r1_dongarra2020report} with A64FX processors optimized oneDNN to scale applications \cite{r2_sato2020co}, while the Sunway Taihulight system \cite{fu2016sunway} employed swCaffe \cite{li2018swcaffe} for distributed training. CPUs have also been used for memory-intensive tasks: Huang \cite{huang2021hierarchical} proposed hierarchical training for recommendation models, and Intel researchers \cite{kalamkar2020optimizing} optimized deep recommendation workloads on Skylake/Cascade Lake CPUs. At large scale, You et al. \cite{you2018imagenet} trained AlexNet in 11 minutes using 2,048 Xeon CPUs with LARS. These works demonstrate the feasibility of CPU-based training, particularly for memory-demanding workloads, but they mainly targeted non-Transformer models that are less representative in today’s large-model landscape, and their optimizations were highly case-specific rather than generalizable.

At the operator level, high-performance math libraries provide essential support. oneDNN \cite{r3_onednn} accelerates deep learning across multiple platforms, while Arm-specific efforts such as autoGEMM \cite{r5_wu2024autogemm}, LBBGEMM \cite{r6_wei2022lbbgemm}, and automatic micro-kernel generation \cite{r4_alaejos2024automatic} optimize GEMM computations through dynamic tiling, auto-tuning, or code generation. Although these methods achieve strong performance on traditional CPUs, they remain focused on vector-based instruction sets and fail to exploit newer architectural features such as matrix acceleration units (MAUs) and on-package high-bandwidth memory, limiting their effectiveness on processors like LX2.

At the framework level, diffusion and Transformer models have motivated specialized optimizations. DiffusionPipe \cite{r10_tian2024diffusionpipe} explored pipeline-parallel training strategies for DiT, while general Transformer frameworks such as xFormers \cite{r8_xFormers2022}, MEATTEN \cite{r11_fu2024optimizing}, OneFlow \cite{r9_yuan2021oneflow}, DeepSpeed \cite{rasley2020deepspeed}, and Megatron \cite{shoeybi2019megatron} introduced operator fusion, modular attention kernels, and distributed scaling methods. Among them, MEATTEN targets Arm CPUs with fusion and tiling optimizations. However, the majority of these frameworks are GPU-centric, making them difficult to port to CPU clusters. This gap highlights the need for CPU-specific frameworks that can exploit new architectural features while supporting representative Transformer-based generative models such as DiT.

\section{Preliminaries and Analysis}
\subsection{Architecture of Diffusion Transformers}
\begin{figure}
    \centering
    \includegraphics[width=1.0\linewidth]{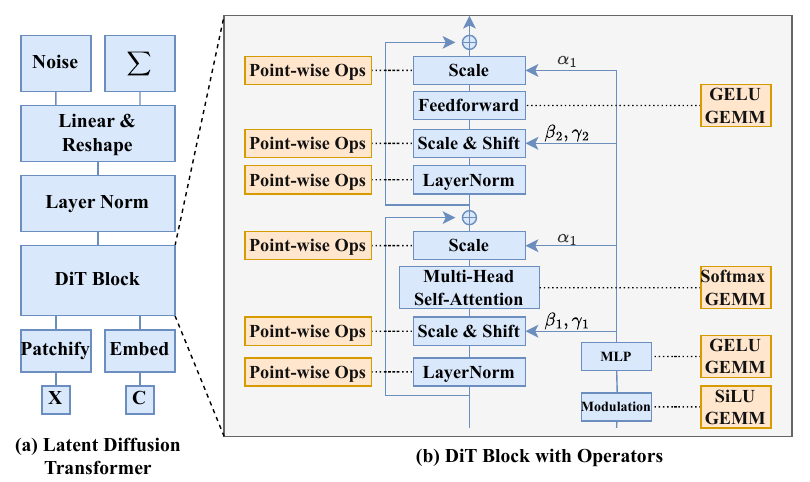}
    \captionsetup{font=small}
    \caption{The workflow of DiT and corresponding operators.}
    \label{fig:DiT}
\end{figure}

Diffusion Transformer (DiT) is a state-of-the-art visual generation framework built upon Denoising Diffusion Probabilistic Models (DDPM). 
By leveraging self-attention mechanisms and modular design, DiT significantly enhances the scalability of the model and the quality of generation. 
DiT revolutionizes noise prediction by architecting a fully transformer based network that processes latent variables through the following components.

\textit{1. Patchify:} the input latent variable $x_t \in \mathbb{R}^{H \times W \times C}$ is divided into $ N = \frac{H}{p} \times \frac{W}{p}$ patches of size $p \times p$, then linearly projected into a $D$-dimensional sequence.
\begin{equation}
    z_t = \text{Linear}(\text{Flatten}(x_t)) \in \mathbb{R}^{N \times D}.
\end{equation}

\textit{2. Conditional Transformer Block:} each block comprises Multi-Head Self-Attention (MSA), Adaptive Layer Normalization (AdaLN) and the Feed-Forward Network (FFN). AdaLN encodes the timestep $t$ and the class label (or other conditions) $y$ into the modulation parameters, which can be formalized as equation~\ref{eq:adaLN}.
  \begin{equation}
    \gamma_t, \beta_t = \text{MLP}(t), \quad \text{AdaLN}(h) = \gamma_t \odot \text{LayerNorm}(h) + \beta_t
    \label{eq:adaLN}.
  \end{equation}

Then the MSA models global contextual relationships based on the attention mechanism.:
  \begin{equation}
    \text{Attention}(Q, K, V) = \text{Softmax}\left( \frac{QK^\top}{\sqrt{D}} \right)V,
  \end{equation}
  where \( Q, K, V \in \mathbb{R}^{N \times D} \) are generated via linear transformations of the input sequence.

\textit{3. Output Reconsturction:} the final sequence is reconstructed into the noise prediction $ \epsilon_\theta(x_t, t)$  via a de-patchify layer.

We have outlined the specific operators involved in the DiT model structure, as illustrated in Figure~\ref{fig:DiT}. During the entire training process, the computational workload is primarily dominated by matrix multiplication kernels (aten::matmul or aten::addmm operators and their corresponding backward processes). Additionally, several commonly used AI operators, such as activation functions (e.g., Gelu, Silu, and Softmax), normalization operations (e.g., LayerNorm), and various mathematical operations (e.g., tanh, scale, and addition), also contribute to the overall computational load.

\subsection{Overall Architecture of LX2 CPU}
\label{sec:arch}

This work is based on the 256-node cluster named LS pilot system. Each node is equipped with two LX2 high-performance CPUs having more than 256 cores. The LX2 CPU is a system on a chip integrating two computing Dies in the package. Each Die is equipped with 128GB off-die DDR memory, along with 4 NUMA domains. Within each NUMA domain, CPU cores share the on-package memory (OPM) and the L2 cache. In order to improve the data movement efficiency between DDR and OPM, System Direct Memory Access (SDMA) interface is implemented in each CPU Die. Dies are also interconnected through an LXLink network, with each network card providing a bidirectional bandwidth of 48 GB/s. In each core, it supports vector and matrix computation based on the vector acceleration units (VAU) and matrix acceleration unit (MAU), with double-precision floating-point SIMD instructions and $8\times8$ matrix computation within the pipeline. The cores operate at a frequency over 1.3GHz.

\begin{figure}
    \centering
    \includegraphics[width=1.0\linewidth]{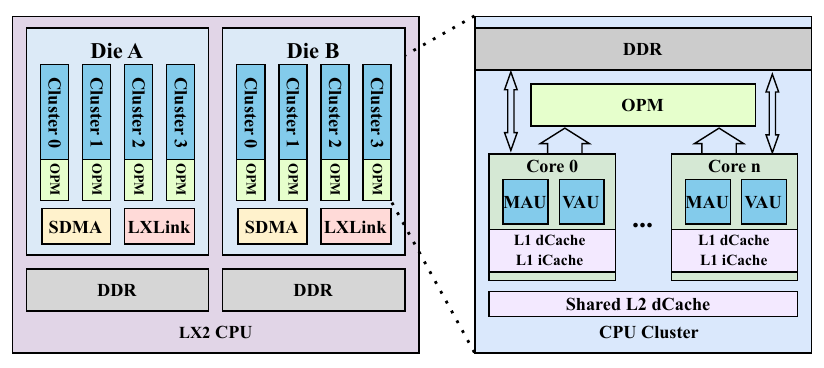}
    \captionsetup{font=small}
    \caption{Architecture of the LX2 CPU.}
    \label{fig:enter-label}
\end{figure}

\subsection{Hierarchical OPM-DDR Hybrid Memory}
\label{sec:bg-arch}
\subsubsection{Limited memory capacity}
In the hierarchical OPM-DDR hybrid memory architecture, the first factor to consider is the memory capacity. While OPM offers excellent bandwidth performance, its capacity is relatively limited, much smaller than that of DDR memory. Furthermore, each CPU cluster is equipped with only 32GB of DDR space, which presents a significant challenge for deep learning training. Current deep learning models typically require tens of gigabytes of memory to store model parameters and training states, making it difficult to scale the model's batch size, thus limiting hardware efficiency during the training process. As the model size increases, insufficient memory not only affects training efficiency but may also lead to memory overflow, further exacerbating the bottleneck in hardware resources.

\subsubsection{Memory access performance} Memory access performance is another critical consideration. The bandwidth of OPM is independent between CPU clusters following the NUMA architecture, which means that remote OPM access across CPU clusters significantly increases latency and reduces bandwidth. Specifically, remote OPM access within the same Die results in about a 50\% decrease in bandwidth compared to its original value; when accessing across Dies within the same CPU, the bandwidth drops to 37\%; and for inter-CPU access within the same node, the bandwidth is reduced to only 10\% of the original value. In contrast, DDR bandwidth is shared within a Die. Although each CPU cluster in the NUMA architecture has its own DDR private space, experimental results show that local and remote DDR access within the same Die exhibit similar performance. When accessing remote DDR across Dies within the same CPU, performance drops to 75\% of its original value, and when accessed across CPUs within the same node, the performance decreases to 37\%, similar to the performance loss observed with remote OPM access.

These performance results suggest that in designing an efficient memory access strategy, it is crucial to minimize remote OPM access across NUMA nodes and DDR access across Dies to avoid performance bottlenecks caused by bandwidth loss. On the other hand, the consistent DDR access performance within a Die offers great opportunity for resource management across NUMA nodes.

\subsection{Matrix Acceleration Units}
The Matrix Acceleration Unit (MAU) in the LX2 CPU supports double-precision 8x8 matrix computation within the pipeline, offering powerful computational capabilities. Maximizing the utilization of the MAU is crucial for performance in deep learning applications. The key factor that determines whether the MAU can deliver optimal performance is whether data can be fed into the vector registers promptly, enabling better pipeline throughput for the MAU.
To figure out this issue, we conducted a series of profiling tests and identified several important factors that affect the efficiency of MAU-based GEMM computations. First, L2 cache utilization plays a critical role—when memory access patterns that are more L2 cache-friendly are used to supply data for matrix computations, MAU utilization improves. Second, the TLB miss ratio is another influential factor. We found that using huge page memory significantly reduces the TLB miss ratio, which in turn enhances MAU utilization, benefiting both OPM and DDR memory. Lastly, memory bandwidth also impacts performance; when matrix data resides in OPM, computation efficiency is higher than when the data is located in DDR.

\label{sec:bg-automem}
To improve MAU utilization, it is essential to adopt memory access patterns that are friendly to L2 cache and to ensure that matrix data is primarily stored in OPM huge page memory. However, these requirements make manual data flow management complex and may result in application code that is difficult to maintain and reuse. Therefore, it is necessary to explore automated methods for efficiently managing data flow across hybrid memory.

\section{Design and Implementations}
\label{sec:design}
\subsection{Parallel Strategy: CFTP with DP}
\begin{figure}
    \centering
    \includegraphics[width=1.0\linewidth]{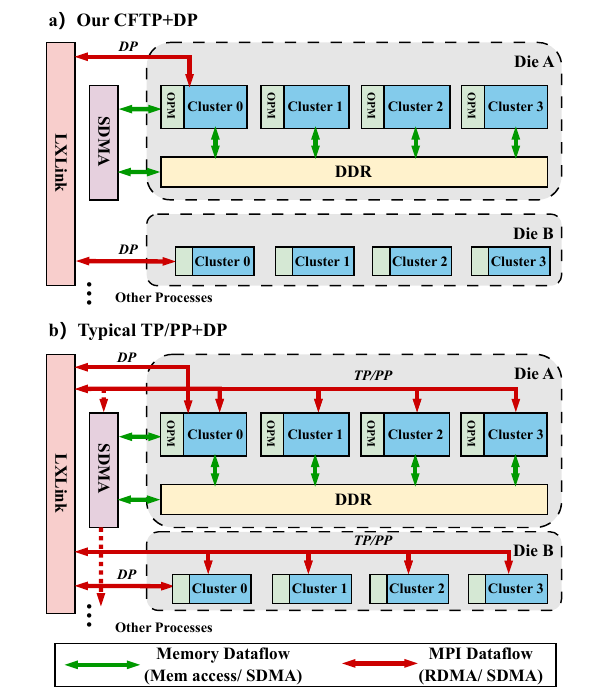}
    \captionsetup{font=small}
    \caption{Communication dataflow comparison between typical TP/PP and our CFTP.}
    \label{fig:CFTP_comm}
\end{figure}

\begin{figure}
    \centering
    \includegraphics[width=1.0\linewidth]{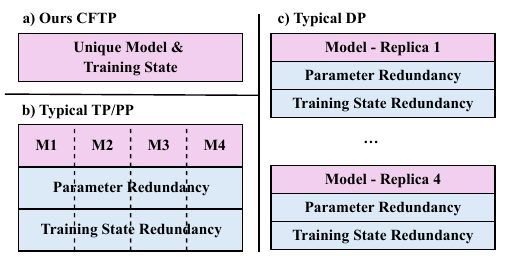}
    \captionsetup{font=small}
    \caption{Memory usage comparison for a Die between typical TP/PP, DP and our CFTP.}
    \label{fig:CFTP_mem}
\end{figure}

\subsubsection{Data Parallel}

Due to the significant efficiency loss when accessing remote OPM and DDR across Dies, we allocate the computational resources of a single Die to each DP process, rather than using a larger resource range. For DP implementation, we follow the standard PyTorch DDP approach, with each die storing a full replica of the model.

\subsubsection{CFTP: Communication-Free Tensor Parallel}

\label{sec:cftp}
As analyzed in Section~\ref{sec:bg-arch}, CPU clusters within a die share the same DDR controller, and DDR access—whether local or remote within the die—exhibits consistent latency and bandwidth. Moreover, DDR bandwidth is substantially higher than that of the LXLink interconnect. Based on these observations, we enable synchronization and sharing among TP nodes via direct memory operations instead of MPI processes. This eliminates the communication overhead of conventional TP methods and allows TP nodes to reside in the same process space, reducing both communication and memory redundancy during model scaling.

Figure~\ref{fig:CFTP_comm} compares the communication overhead of different die-level parallelism methods. In Figure~\ref{fig:CFTP_comm}(a), our method assigns one process per die, with TP implemented across CPU clusters within the die. Training requires MPI only for gradient reduction across dies, resulting in just four MPI processes per compute node. By contrast, Figure~\ref{fig:CFTP_comm}(b) shows typical TP or PP implementations. These approaches require not only gradient synchronization between DPs but also synchronization or activation transfers within each node. While SDMA can reduce LXLink traffic, it introduces contention with DDR–OPM transfers, degrading memory performance. Unlike MPI in-place operations, SDMA also requires additional memory buffers to ensure reliability, causing extra memory overhead and frequent allocations. Consequently, typical TP/PP schemes increase the MPI process count to 16 per node, complicating the overlap of memory and communication flows and hindering balanced hardware utilization.

For memory usage within a single die (Figure~\ref{fig:CFTP_mem}), our method in Figure~\ref{fig:CFTP_mem}(a) manages all memory within one process, enabling flexible partitioning of any tensor (inputs, weights, gradients, activations) during matrix computations. This eliminates the need to exchange intermediate results, making CPU clusters fully independent and requiring only a single copy of the model and training state. By contrast, typical TP/PP methods increase memory consumption due to cross-process partitioning. TP requires partitioning to be fixed at initialization, usually limited to weights, leading to broadcast or aggregation of inputs and activations. PP caches intermediate activations between stages, with memory usage growing alongside pipeline depth, and may introduce load imbalance across clusters. Similarly, DP stores full model replicas on each device, which results in high memory redundancy and becomes a bottleneck to batch size scalability as device count grows.

In summary, the combined use of CFTP and DP significantly alleviates communication and memory bottlenecks under hierarchical memory architectures.

\subsection{AutoMem: Automatic Memory Dataflow Management Module}

\begin{algorithm}[t]
\caption{Our AutoMem\_Module Wrapper}
\SetKwFunction{Forward}{forward($X_i$)}
\SetKwFunction{Backward}{backward($dX_{i+1}$)}
\SetKwFunction{Train}{train()}
\SetKwFunction{Preforward}{pre\_forward()}
\SetKwFunction{Postforward}{post\_forward()}
\SetKwFunction{Prebackward}{pre\_backward()}
\SetKwFunction{Postbackward}{post\_backward()}
\SetKwFunction{Warmupforward}{warmup\_forward()}
\SetKwFunction{ClassAutomem}{\textbf{Class} AutoMem\_Module(nn.Module)}
\SetKwProg{Def}{def}{:}{}
\SetKwProg{Method}{}{:}{}

\BlankLine
\Method{\ClassAutomem}{
    Ori\_Module = nn.Module\;

    \Def{\Forward}{
        \Method{\Preforward}{
            Barrier checking whether $X_i$, $W_i$ has been loaded\;
            Prefetch $W_{i+1}$\;
            Register post\_backward() hook on $W_i.\texttt{acc\_grad}$\;
            set MemType(OPM-HugePage)\;
        }
        $X_{i+1} \gets \texttt{Ori\_Module.forward}(X_i, W_i)$\;

        \Method{\Postforward}{
            Offload $X_i$, $W_i$\;
            Register pre\_backward() hook on $X_{i+1}$\;
            set MemType(DDR)\;
        }
    }

    \Def{\Backward}{
        \Method{\Prebackward}{
            Barrier checking $dX_{i+1}$, $X_i$, $W_i$\;
            Prefetch $W_{i-1}$, $X_{i-1}$\;
            set MemType(OPM-HugePage)\;
        }
        $dX_i, dW_i \gets \texttt{Ori\_Module.backward}()$\;

        \Method{\Postbackward}{
            Offload $W_i$, $X_i$, $dW_i$\;
            set MemType(DDR)\;
        }
    }
}

\Def{\Train}{
    \Method{\Warmupforward}{
        Record execution order of Modules\;
        Initialize\_MemPool()\;
        Activation\_Reference\_track()\;
    }
    \texttt{AutoMem\_Module.forward()}\;
    \ldots
}

\end{algorithm}

\begin{figure*}[h!]
    \centering
    \includegraphics[width=1.0\textwidth]{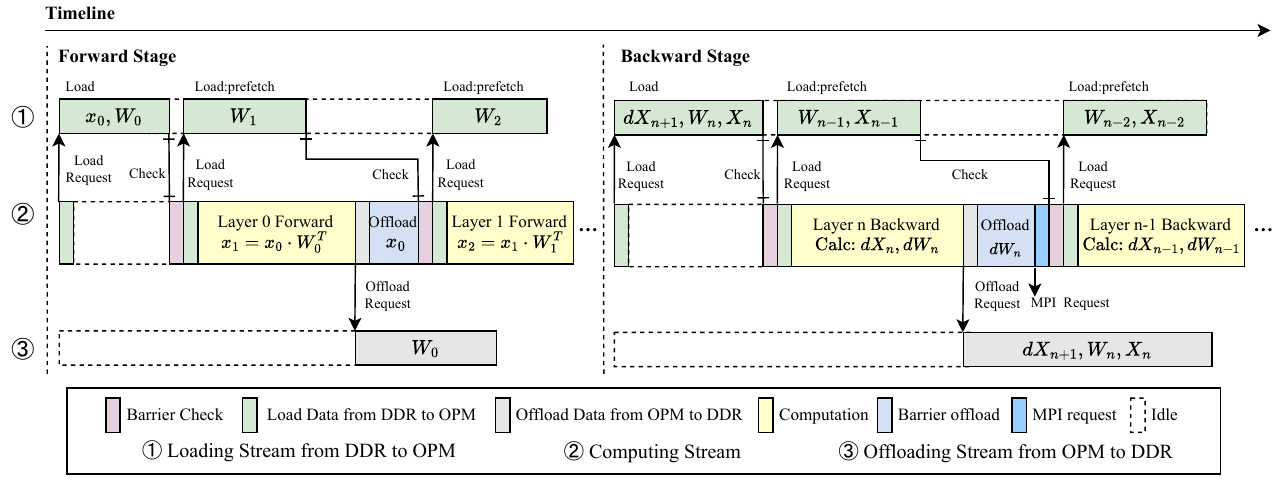} 
    
    \captionsetup{font=small}
    \caption{Overlapped dataflow based on our AutoMem. Different stream are distributed to different CPU cores.}
    \label{fig:automem}
\end{figure*}

Optimizing the performance of HPC-oriented CPUs with OPM-DDR hierarchical memory requires fine-grained management of dataflow across memory levels. Manual strategies are complex and hard to generalize; therefore, we design AutoMem, a non-intrusive automated memory management module for deep learning training. AutoMem improves utilization of high-bandwidth on-package memory (OPM), enhances operator efficiency, and accelerates end-to-end DiT training—all without modifying model-level code—while ensuring that optimization strategies remain reusable for future workloads.

Algorithm 1 outlines the implementation of AutoMem. The original \texttt{nn.Module} is wrapped by \texttt{AutoMem\_Module}, with \texttt{pre\_forward()} invoked before the forward pass and \texttt{post\_forward()} after it. In \texttt{pre\_forward()}, AutoMem checks whether inputs and weights ${X_i, W_i}$ reside on OPM; if not, they are loaded. It also prefetches the next layer’s weights $W_{i+1}$. A \texttt{post\_backward()} hook is registered on the \texttt{AccumulateGrad} object of $W_i$, triggered once gradients are accumulated. Memory allocation is temporarily switched to OPM HugePages to ensure forward-context data remain on OPM. The forward computation then produces $X_{i+1}$. In \texttt{post\_forward()}, $X_i$ and $W_i$ are offloaded back to DDR, and a hook is set on $X_{i+1}$ to trigger its gradient loading. Finally, allocation defaults back to DDR to avoid overusing OPM HugePages.

The backward process follows a conceptually similar design, with \texttt{pre\_backward()} and \texttt{post\_backward()} hooks. \texttt{pre\_backward()} ensures required tensors ($dX_{i+1}, W_i, X_i$) are loaded into OPM and prefetches $W_{i-1}$ and $X_i$ for the next backward step. Memory allocation mode is switched to OPM HugePages, then the original backward pass produces $dX_i$ and $dW_i$. In \texttt{post\_backward()}, $W_i$ and $X_i$ are offloaded to DDR and allocation is reset. This completes one forward and backward cycle.

During training initialization, AutoMem performs a warm-up forward pass to record module dependencies and enable prefetching. It then initializes a pinned DDR HugePage memory pool to ensure sufficient and stable bandwidth for offloading operations. This lightweight pinned memory pool, integrated as a core component of AutoMem, provides efficient support for memory transfers. In addition, AutoMem tracks the reference relationships of inputs and activations at the memory level during the warm-up stage to determine when they can be safely offloaded during training.

All load and offload operations are implemented with SDMA and managed by dedicated CPU cores (Figure~\ref{fig:automem}). Stream 2 serves as the main computation flow, occupying nearly the entire die. In contrast, Streams 1 and 3 are dedicated to data transfers, each using only one independent CPU cores. At the beginning of forward and backward passes, the first layer blocks until data are loaded into OPM; afterwards, AutoMem overlaps asynchronous prefetching and offloading with computation. For gradient tensors requiring global communication, blocking offloading is performed on the main computation cores to ensure correctness, as indicated by the blue blocks in Figure~\ref{fig:automem}. This design achieves efficient overlap of computation and memory operations while incurring minimal additional core overhead via independent SDMA devices.

\subsection{Operator-Level Optimization}

\begin{figure*}[htbp]
    \centering
    \includegraphics[width=1.0\textwidth]{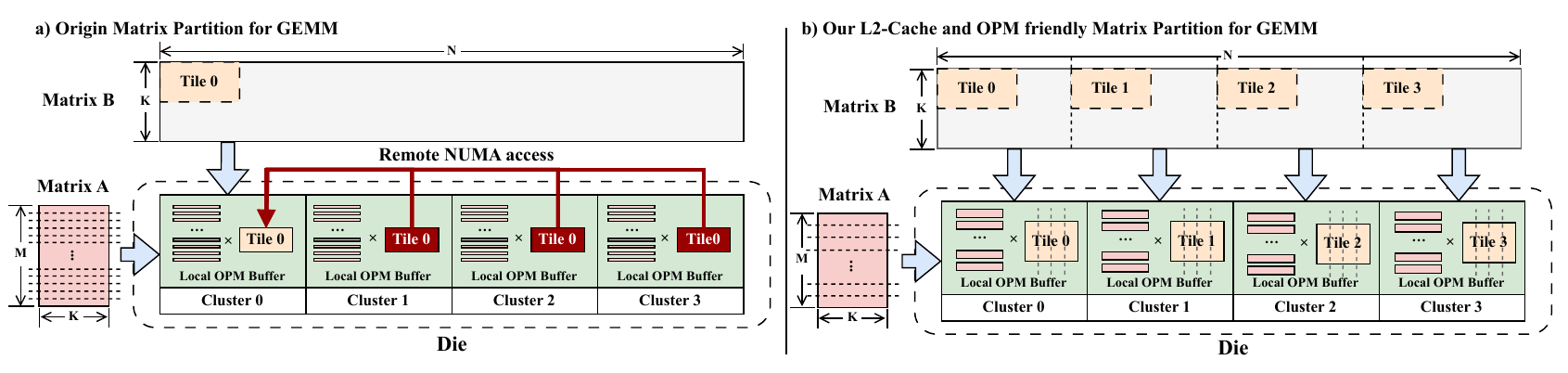} 
    
    \captionsetup{font=small}
    \caption{The diagram of our optimized GEMM with L2 Cache and memory access-friendly tiling strategy.}
    \label{fig:gemm}
\end{figure*}

To improve the utilization of the Matrix Acceleration Unit (MAU) and Vector Acceleration Unit (VAU), we optimized computationally intensive components of the model, including GEMM and commonly used AI operators. For GEMM, we redesigned the tiling strategy to increase L2 cache efficiency and balance thread workloads. In addition, we optimized operators such as GeLU, SiLU, LayerNorm, Softmax, and FlashAttention. Finally, building on these operator-level improvements, we introduced system-level tuning strategies—including parallel partitioning, NUMA-aware data transfer, and multi-level buffering—to further enhance scalability. Together, these optimizations are integrated into the HCOps extension package, enhancing both the PyTorch framework and the native software stack.

\subsubsection{GEMM: numa and L2 cache friendly matrix tiling}

We adopt the vendor-provided BLAS library in the native software stack, referred to as nativeBLAS, as the baseline for GEMM computations. NativeBLAS employs a closed-source kernel with instruction reordering and tiling, where matrix $B$ is shared and matrix $A$ is partitioned along the $M$-dimension. Specifically, the main thread loads a tile of $B$ into the local OPM buffer, performs packing to ensure memory access continuity, and then shares the tile with all threads. Each thread $i$ loads its assigned sub-matrix $A_i$ into a private buffer, multiplies it with $B_{Tile}$, and writes the result to $C_i$.

This design, however, suffers from two key drawbacks. First, except for the NUMA node where the main thread resides, other clusters must fetch $B_{Tile}$ from remote OPM, which introduces high latency. Second, within each cluster, a large number of cores simultaneously access different memory addresses of matrix $A$. These accesses converge on the shared L2 cache, creating severe contention. As a result, the effective L2 cache hit ratio drops significantly, preventing data reuse and ultimately constraining the computational throughput of the MAU.

To address these issues, we propose a NUMA- and cache-aware 2D partitioning scheme (Figure~\ref{fig:gemm}(b)). At the outer level, matrix $B$ is divided along the $N$-dimension into sub-matrices assigned to CPU clusters (e.g., four partitions for four clusters). Each cluster independently computes $A \times B_{cid}$, with its tiles packed into local OPM buffers and retained in the L2 cache using loop unrolling. At the inner level, $A_{Tile}$ is partitioned into $m$ segments along $M$ and $B_{Tile}$ into $n$ segments along $N$, resulting in $m \times n$ subtasks distributed evenly across threads. Compared with nativeBLAS, which partitions $A_{Tile}$ into $m_{\text{native}}$ segments, our approach reduces the partition count to $m$ ($m < m_{\text{native}}$), thereby improving cache reuse. This significantly increases L2 cache utilization and enhances the computational efficiency of the MAU.

\subsubsection{Optimization of AI operators}

Beyond GEMM, we optimized several key operators on LX2 to exploit the VAU and MAU more effectively, including GeLU, SiLU, Softmax, LayerNorm, FlashAttention, and fundamental mathematical kernels. For example, as for Tanh-GELU, a hybrid approximation scheme accelerates forward and backward passes by $13.3\times$ and $12.9\times$ with negligible loss of accuracy. For the AdamW optimizer, an operator-fusion design reduces memory writes and achieves a $12.5\times$ iteration speedup. In addition, recomputation strategies for FlashAttention and fused multiply-add kernels reduce memory residency and overall memory demand. These optimizations, integrated into HCOps, complement the GEMM improvements and contribute to end-to-end performance gains.

\subsubsection{Scalability-oriented Tuning}
To further address scalability bottlenecks introduced by multi-core execution and hierarchical memory, we performed systematic tuning of GEMM parallelization and synchronization strategies. The optimizations combined NUMA-aware data transfer, asynchronous inter-core communication, packing, double/triple buffering for memory and registers, tiling, and thread reuse. By exploring parameter combinations across these techniques, we identified the best-performing implementation, which forms the Tuned version of our framework and delivers the highest single-node performance.

\subsection{Asynchronous Communication Backends}
We implement a customized MPI-based backend to address the limitations of PyTorch’s native MPI backend and to achieve improved overlap between communication and computation. The native backend does not support asynchronous collectives such as \texttt{MPI\_Iallreduce}. Moreover, the worker threads responsible for communication are created only after PyTorch DDP initialization and are managed uniformly by PyTorch’s native OpenMP runtime. As a result, communication requests and computation tasks are scheduled on the same cores, forcing frequent context switches between computation and communication. This degrades both communication throughput and computational efficiency.
In addition, communication and computation tasks each introduce their own synchronization mechanisms. When executed on the same cores, the contention for hardware resources further exacerbates performance loss. By customizing the MPI backend, we enable the use of asynchronous communication interfaces and allow explicit binding of communication and computation tasks to different CPU cores. This design reduces resource contention and achieves more effective overlap between communication and computation.

\section{Experiments}

\subsection{Setting Details}

\textbf{Hardware and Software.}
All optimizations in this study are conducted on the LX2 CPU hardware architecture within the LS Pilot System, which consists of 256 nodes (detailed in Section~\ref{sec:arch}). The software environment is based on the OpenEuler 22.03 operating system, with Clang 17.0.0 as the compiler and OpenMPI for communication. For mathematical and deep learning acceleration libraries, a combined baseline of nativeBLAS optimized for the LX2 CPU together with oneDNN 3.6.2 is employed. The model implementation and optimizations are developed on PyTorch 2.5.1. To ensure the correctness of results, additional validation experiments are performed on the NVIDIA H100 SXM5 GPUs equipped with 80 GB of HBM3 memory.

\textbf{Model and Dataset}
The proposed DiT-HC architecture is evaluated across multiple model scales, including DiT-S/2, DiT-B/2, DiT-L/2, and DiT-XL/2, to demonstrate the effectiveness of our optimizations under different model sizes and kernel configurations. Training settings follow the original DiT implementation, employing the AdamW optimizer with a base learning rate of $10^{-4}$.

Experimental evaluation involves three complementary datasets representing different modalities: the general-purpose ImageNet dataset, and two remote sensing datasets derived from Gaofen-2 and Sentinel-2 satellites. The Gaofen-2 dataset provides 4-band multispectral imagery at spatial resolution of 2m/pixel, while the Sentinel-2 dataset contains 13-band multispectral imagery at a spatial resolution of 10m/pixel.

\textbf{Evaluation.} Training is performed with the Mean Squared Error (MSE) loss between predicted and target latents, enforcing pixel-level reconstruction fidelity for both training from scratch and fine-tuning. For final evaluation, we adopt the Fréchet Inception Distance (FID) to assess the perceptual realism of generated images. FID is computed as the 2-Wasserstein distance between the feature distributions of real and generated images, where the mean vector and covariance matrix are estimated from features extracted using a pretrained Inception-v3 network. Lower FID values indicate that the generated distribution is closer to the real data distribution.

\subsection{Accuracy Validation}

To verify the correctness of our optimized training methods, we designed two types of experiments: an early-stage training scenario, where the model is trained from scratch on ImageNet, and a late-stage training scenario, where a well-converged pretrained model is fine-tuned on new remote sensing datasets.

\begin{figure}
    \centering
    \includegraphics[width=1.0\linewidth]{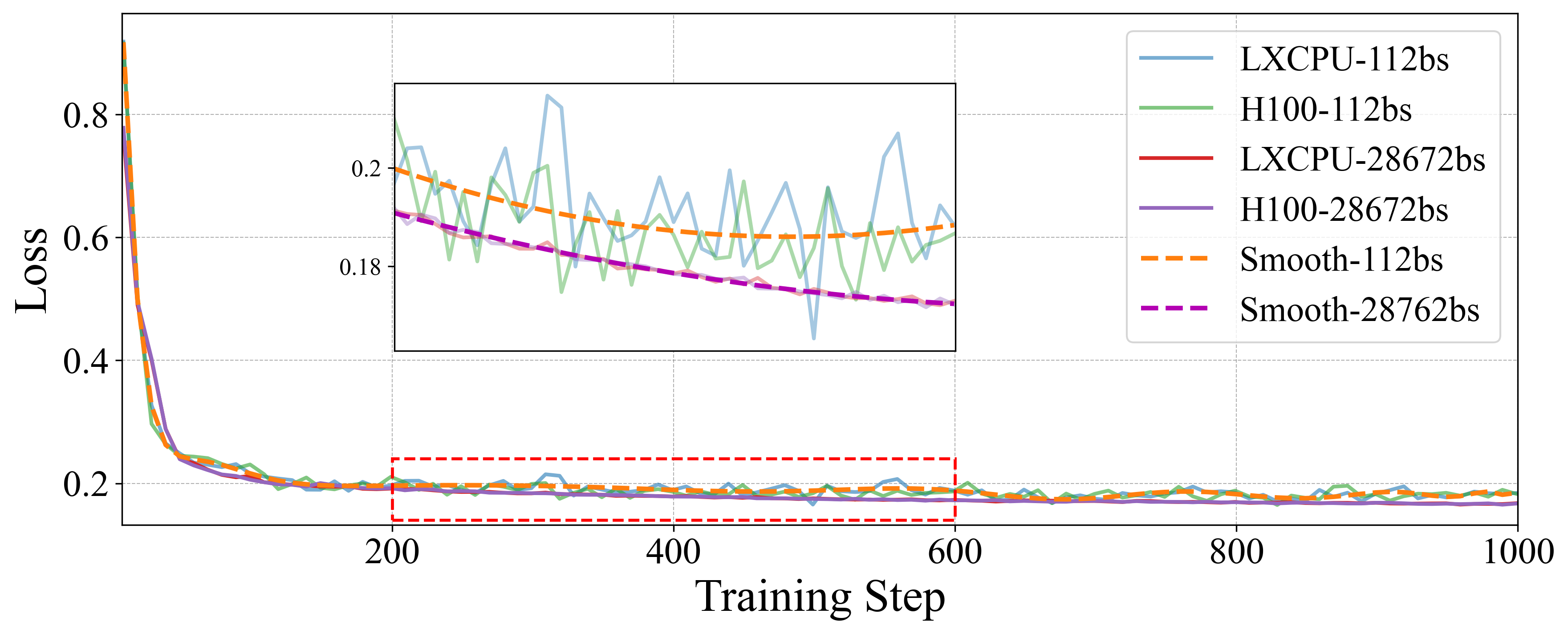}
    \captionsetup{font=small}
    \caption{Comparison of loss curve of pretraining on ImageNet.}
    \label{fig:train_loss}
\end{figure}

In the training-from-scratch experiments, we conducte 1,000 training steps under two batch-size settings: a single-node batch size of 112 and the largest global batch size in our experiments, 28672. This design serves two purposes: validating the consistency of results between the LX2 CPU cluster and the NVIDIA H100 GPU, 
In the training-from-scratch stage, we ran DiT-XL/2 on ImageNet for 1,000 steps with two batch-size settings: a single-node batch size of 112 and the largest global batch size in our experiments, 28,672. This setup verifies the consistency of results between the LX2 CPU cluster and the NVIDIA H100 GPU, and demonstrate that even at the largest training scale, the model can still converge effectively. In Figure~\ref{fig:train_loss}, different colors denote the per-step loss values obtained on different platforms and batch-size settings. The two dashed lines show smoothed versions of the oscillating loss curves, providing a clearer view of the convergence trend.

\begin{figure}
    \centering
    \includegraphics[width=0.9\linewidth]{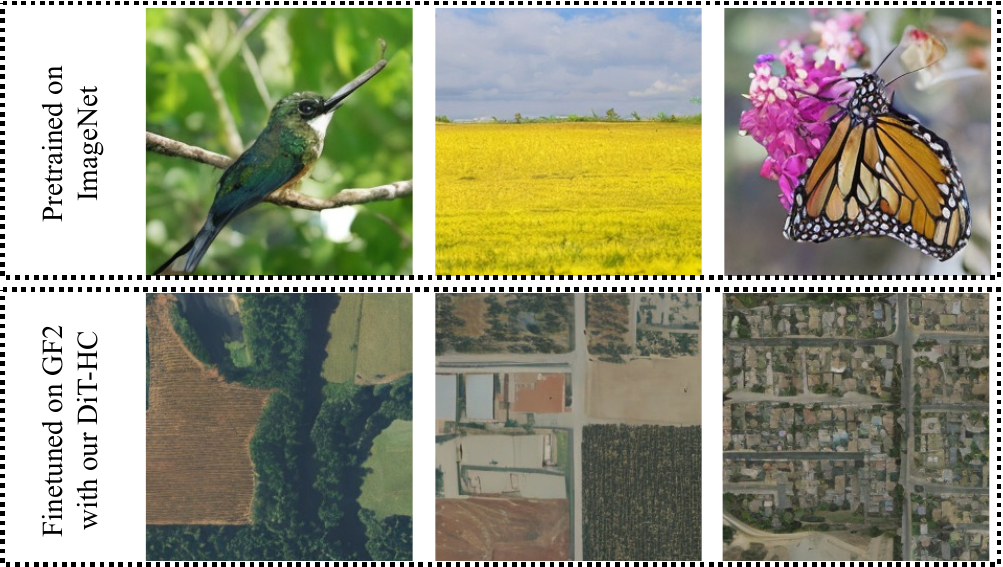}
    \captionsetup{font=small}
    \caption{Visual results of the original pretrained model and fine-tuned model on Gaofen-2 by our DiT-HC.}
    \label{fig:visual}
\end{figure}

In the fine-tuning stage, the pretrained model was adapted to the Gaofen-2 and Sentinel-2 datasets for 1,000 steps using two batch-size settings: 112 and 28,672. For each configuration, 10,000 images were generated and compared with 10,000 real samples from the corresponding dataset to compute the Fréchet Inception Distance (FID). Table~\ref{tab:acc} summarizes the results: with a batch size of 112, the proposed optimized training method enables stable DiT training on LX2 while maintaining generation quality comparable to the NVIDIA H100. With a batch size of 28,672, the results demonstrate that the model can continue learning effectively even under extremely large-scale training. Figure~\ref{fig:visual} presents visual examples produced by the fine-tuned DiT-HC model, further confirming the reliability of our framework and its ability to generalize across different domains of scientific data.

\begin{table}[t]
  \captionsetup{font=small}
  \caption{FID ($\downarrow$ is better) results on remote sensing datasets, where bs = 112 compares GPU and LX2; bs = 28{,}672 shows effectiveness.}
  \label{tab:acc}
  \footnotesize
  \centering
  \begin{tabular}{l|ccc}
    \toprule
    Dataset & GPU (112) & LX2 (112) & LX2 (28{,}672) \\
    \midrule
    Gaofen\textendash2   & 30.44 & 28.51 & 24.70 \\
    Sentinel\textendash2 & 24.97 & 29.03 & 22.61 \\
    \bottomrule
  \end{tabular}
\end{table}

\subsection{Results for Different Model Size}

\begin{table}
  \centering
  \captionsetup{font=small}
  \caption{Comparison of single-step execution time and peak memory consumption across different strategies.}
  \footnotesize
    \begin{tabular}{c|c|ccc|ccc}
    \toprule
    \multirow{2}{*}{\parbox{0.7cm}{\centering Batch \\Size}} & \multirow{2}{*}{\parbox{0.9cm}{\centering Model\\Size}} & \multicolumn{3}{c|}{Iteration Time (s)} & \multicolumn{3}{c}{Peak Memory (GB)} \\ \cline{3-8} & & \raisebox{-0.3ex}{DP+TP} & \raisebox{-0.3ex}{DP} & \raisebox{-0.3ex}{CFTP} & \raisebox{-0.3ex}{DP+TP} & \raisebox{-0.3ex}{DP} & \raisebox{-0.3ex}{CFTP}
    \\
    \midrule
    \multirow{4}[2]{*}{112} & XL/2  & -     & -     & 13.18& -     & -     & 358 \\
          & L/2   & -     & 23.31 & 9.92& -     & 474   & 304 \\
          & B/2   & 8.7   & 6.55  & 3.97& 309   & 247   & 198 \\
          & S/2   & 3.91  & 2.11  & 2.48& 224   & 182   & 167 \\
    \midrule
    \multirow{4}[2]{*}{224} & XL/2   & -     & -     & 29.15& -     & -     & 346 \\
          & L/2   & -     & -     & 22.88& -     & -     & 290 \\
          & B/2   & 17.89 & 11.45 & 9.75& 270   & 277   & 186 \\
          & S/2  & 6.3   & 3.56  & 3.83& 287   & 197   & 180 \\
    \bottomrule
    \end{tabular}%

  \label{tab:addlabel}%
    {\footnotesize\RaggedRight 
        -: Out Of Memory  \\
    }
\end{table}%

We conduct experiments on DiT models of varying scales, including DiT-S/2, DiT-B/2, DiT-L/2, and DiT-XL/2. The results under different parallel strategies with two global batch sizes are summarized in Table~\ref{tab:addlabel}.
For the \textbf{DP+TP} configuration, the input data is replicated across clusters within each TP group. In the \textbf{DP} setting, each cluster stores a complete copy of the model parameters, gradients, and optimizer states. Both strategies can cause large models to exceed memory capacity. In terms of peak memory usage, \textbf{CFTP} achieves the lowest consumption among all tested cases.
We also observe that \textbf{DP} yields shorter iteration times for smaller models. This is because DP avoids model partitioning by maintaining full copies of the parameters, thereby improving GEMM efficiency at the expense of memory usage.
Although the \textbf{DP+TP} strategy reduces memory consumption by grouping four clusters into a single TP group, it still incurs unavoidable intra-die communication overhead. In contrast, our proposed \textbf{CFTP} eliminates die-level communication among clusters introduced by TP, thereby achieving higher efficiency than DP+TP. Experimental results across all four model scales confirm the effectiveness and general applicability of this strategy.

\subsection{Stepwise Results}
\begin{figure}
    \centering
    \includegraphics[width=1.0\linewidth]{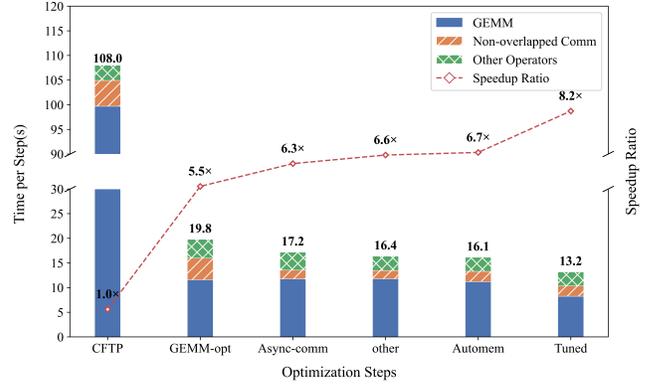}
    \captionsetup{font=small}
    \caption{Stepwise optimization evaluation of our proposed methods.}
    \label{fig:stepwize}
\end{figure}

To evaluate the optimizations proposed in Section~\ref{sec:design}, we perform stepwise ablation experiments.
The baseline training framework is PyTorch compiled with NativeBLAS as the math library, oneDNN as the deep learning acceleration library, and MPI as the communication backend. As discussed in Section~\ref{sec:cftp}, CFTP is included in the baseline to mitigate memory footprint issues.

Building on this baseline, we first introduce GEMM optimization, which raises the speedup to 5.5$\times$. Compared with NativeBLAS, the optimized GEMM reduces cross-NUMA memory access and improves L2 cache hit rates through cache-friendly task partitioning. We then incorporate an asynchronous communication backend; by assigning communication tasks to dedicated cores, the speedup further increases to 6.3$\times$. Adding AI operator optimizations brings an additional improvement, raising the speedup to 6.6$\times$. On top of this, integrating AutoMem—which analyzes the computation graph and prefetches data for both forward and backward passes—slightly improves the performance to 6.7$\times$. Finally, by applying tuning strategies targeting multi-core parallelism and hierarchical memory, including NUMA-aware partitioning, multi-level buffering for memory and registers, and parameter combination optimizations, the single-node speedup ultimately reaches 8.2$\times$.

\subsection{GEMM Optimization}

In DiT training, matrix multiplication constitutes the dominant computational workload, making GEMM optimization a decisive factor for overall performance. In this section, we examine the performance of GEMM operators and compare our implementations against OpenBLAS, NativeBLAS, and oneDNN across different DiT model sizes.
As shown in Table~\ref{tab:gemm_compare}, OpenBLAS provides unsatisfactory performance due to the lack of architecture-specific optimizations for the new processors. Although NativeBLAS and oneDNN are the default libraries, they are not NUMA-aware at the Die level and exhibit low cache hit rates.

To address these issues, we first design our GEMM kernels that maximize data locality and avoid cross-NUMA memory accesses. As shown in Table~\ref{tab:gemm_linear}, these kernels achieve up to 35.49$\times$ speedup over NativeBLAS at the module level.
Building on this foundation, we further introduce our Tuned kernels, which apply a set of refined strategies including NUMA-aware task partitioning, multi-level buffering for memory and registers, and parameter tuning for multi-core parallelism. With these enhancements, our Tuned version achieves even higher performance, further extending the gains obtained by our GEMM kernels.

\begin{table}[htbp]
  \centering
    \captionsetup{font=small}
  \caption{Single-step training time (in seconds) of DiT models using OpenBLAS, NativeBLAS, oneDNN, our GEMM kernels, and our Tuned version. Our GEMM provides a substantial reduction in iteration time compared with standard libraries, while our Tuned version further improves performance through multi-core and memory hierarchy optimizations.}
  \footnotesize
  \setlength{\tabcolsep}{3pt}
    \begin{tabular}{c|ccccc}
    \toprule
    \multirow{2}{*}{\parbox{0.7cm}{\centering Model \\Size}} & \multicolumn{5}{c}{Iteration Time (s)} \\
    \cline{2-6}
     & \raisebox{-0.3ex}{\parbox{1.2cm}{OpenBLAS}} 
     & \raisebox{-0.3ex}{\parbox{1.3cm}{NativeBLAS}} 
     & \raisebox{-0.3ex}{\parbox{1cm}{oneDNN}} 
     & \raisebox{-0.3ex}{our GEMM} 
     & \raisebox{-0.3ex}{our Tuned} \\
    \hline
    XL/2 & 1156.37 & 154.41 & 147.18 & 16.14 & 13.18 \\
    L/2  & 636.08  & 91.72  & 64.86  & 14.52 &  9.92 \\
    B/2  & 181.66  & 41.42  & 19.82  &  5.83 &  3.97 \\
    S/2  &  47.90  & 22.35  &  8.96  &  4.02 &  2.48 \\
    \bottomrule
    \end{tabular}%
  \label{tab:gemm_compare}%
  \vspace{-0.3cm}
\end{table}%

\begin{table}[htbp]
  \centering
  \captionsetup{font=small}
  \caption{Speedup of our optimized GEMM in latency of the individual linear modules compared to nativeBLAS.}
  \label{tab:gemm_linear}

  \setlength{\heavyrulewidth}{1.2pt}
  \setlength{\lightrulewidth}{0.8pt}
\footnotesize
  \begin{tabular}{llr}
    \toprule
    Name & Dimensions & Speedup ($\times$) \\
    \midrule
    qkv\_proj       & $1152 \times 3456$ & 30.10 \\
    o\_proj         & $1152 \times 1152$ & 13.89 \\
    up\_proj        & $1152 \times 4608$ & 16.49 \\
    down\_proj      & $4608 \times 1152$ & 35.49 \\
    condition\_proj & $1152 \times 6912$ & 1.73  \\
    \bottomrule
  \end{tabular}
  \vspace{-0.3cm}
\end{table}

\subsection{Performance Comparison with GPU System}

To position the performance of our CPU-based system relative to GPUs, we conducted a comparative experiment under comparable system-level compute capability. The GPU side consisted of two 8-GPU H100 servers, each H100 delivering 495 TFLOPS in TF32 without sparsity, while the CPU system is built upon HPC-oriented processors whose peak floating-point capability is slightly lower than that of H100 GPUs. 
The two platforms differ markedly in memory and interconnect characteristics. Each H100 server provides ultra-high-bandwidth HBM, NVLink-based intra-node interconnect, and multi-rail InfiniBand for inter-node communication, whereas the CPU platform offers substantially lower aggregated memory bandwidth and inter-processor interconnect bandwidth, resulting in different communication and memory-access behaviors.

With a global batch size of 3{,}584, the CPU system achieves 182.47 TFLOPS sustained performance and an end-to-end training time of 13.5\,s per iteration, compared to 7.6\,s on the H100 system. Although the CPU implementation lags behind GPUs in raw throughput, the results demonstrate the feasibility of scaling generative model training on HPC-oriented CPUs and provide practical insights into architecture-specific optimization.

\subsection{Scalability}

We evaluate scalability under inter-node data parallelism from 1 to 256 nodes. Metrics include average runtime per iteration, achieved FLOPS, and parallel efficiency.

\begin{figure}
    \centering
    \includegraphics[width=1.0\linewidth]{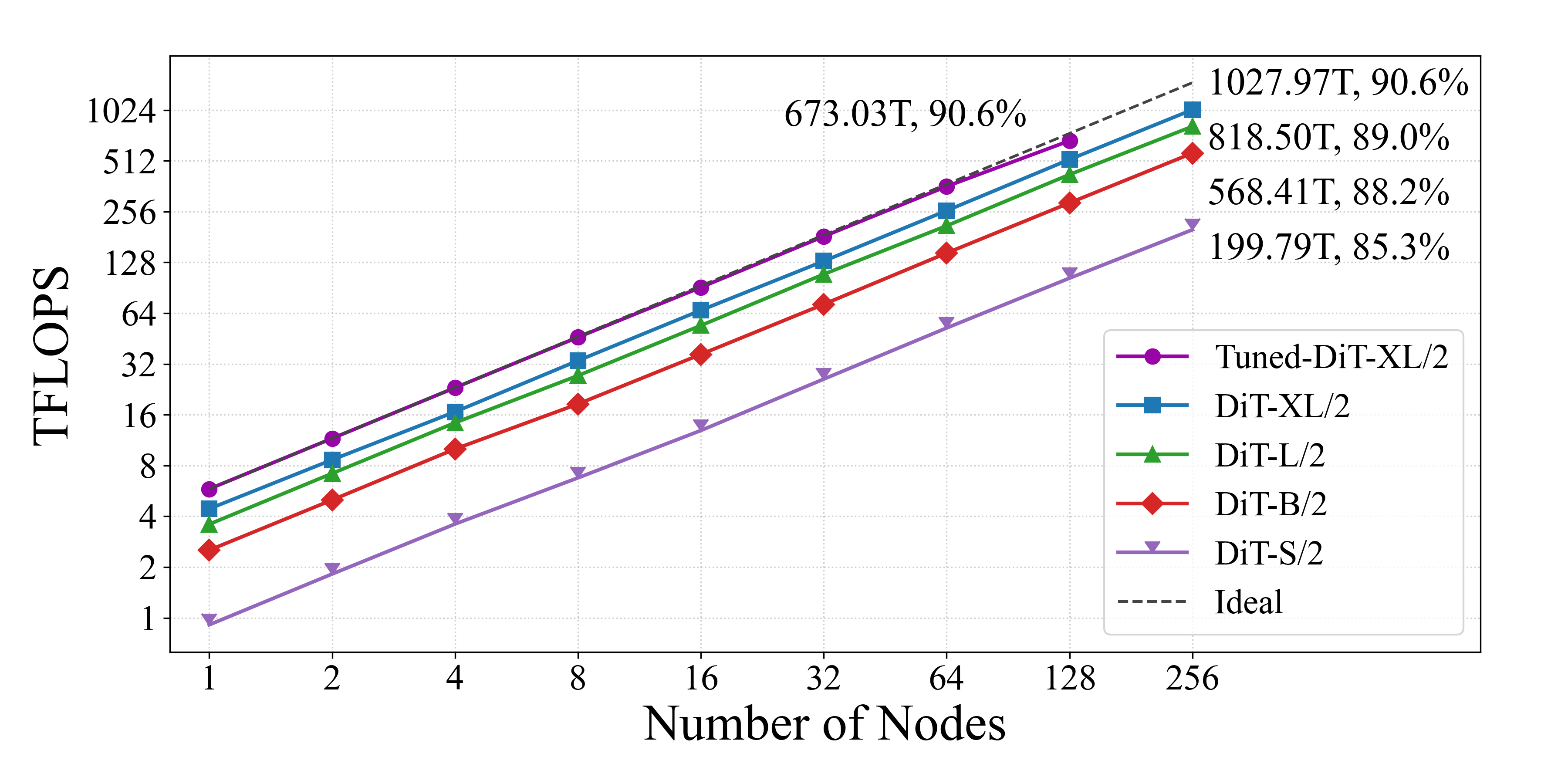}
    \captionsetup{font=small, skip=4pt}
    \caption{Weak scaling results for DiT training.}
    \label{fig:weak_scale}
    \vspace{-0.5cm}
\end{figure}

For weak scaling, the batch size per node is fixed at 112 and the global batch size grows proportionally with the node count. As shown in Figure~\ref{fig:weak_scale}, both the GEMM-optimized and the further tuned configurations exhibit near-linear scaling across four DiT model sizes. Performance and parallel efficiency improve with model size due to higher compute density. On DiT-XL/2, the GEMM-optimized configuration reaches over 1 PFLOPS (FP32) with 90.6\% parallel efficiency at 256 nodes with 1024 processes.

Two factors underpin this efficiency. First, the proposed CFTP strategy reduces the number of ranks and turns intra-die communication into shared-memory accesses, minimizing communication overhead. Second, offloading communication to dedicated cores prevents the compute cores from incurring context switches and synchronization caused by blocked communication.
We also report the results for the further tuned configuration, which adds multi-core and memory hierarchy optimizations on top of the GEMM improvements. Due to machine-allocation constraints, these runs were conducted up to 128 nodes. Within this range, the tuned configuration consistently achieves the best performance, and its scalability trend aligns with other configurations.

\begin{figure}
    \centering
    \includegraphics[width=0.9\linewidth]{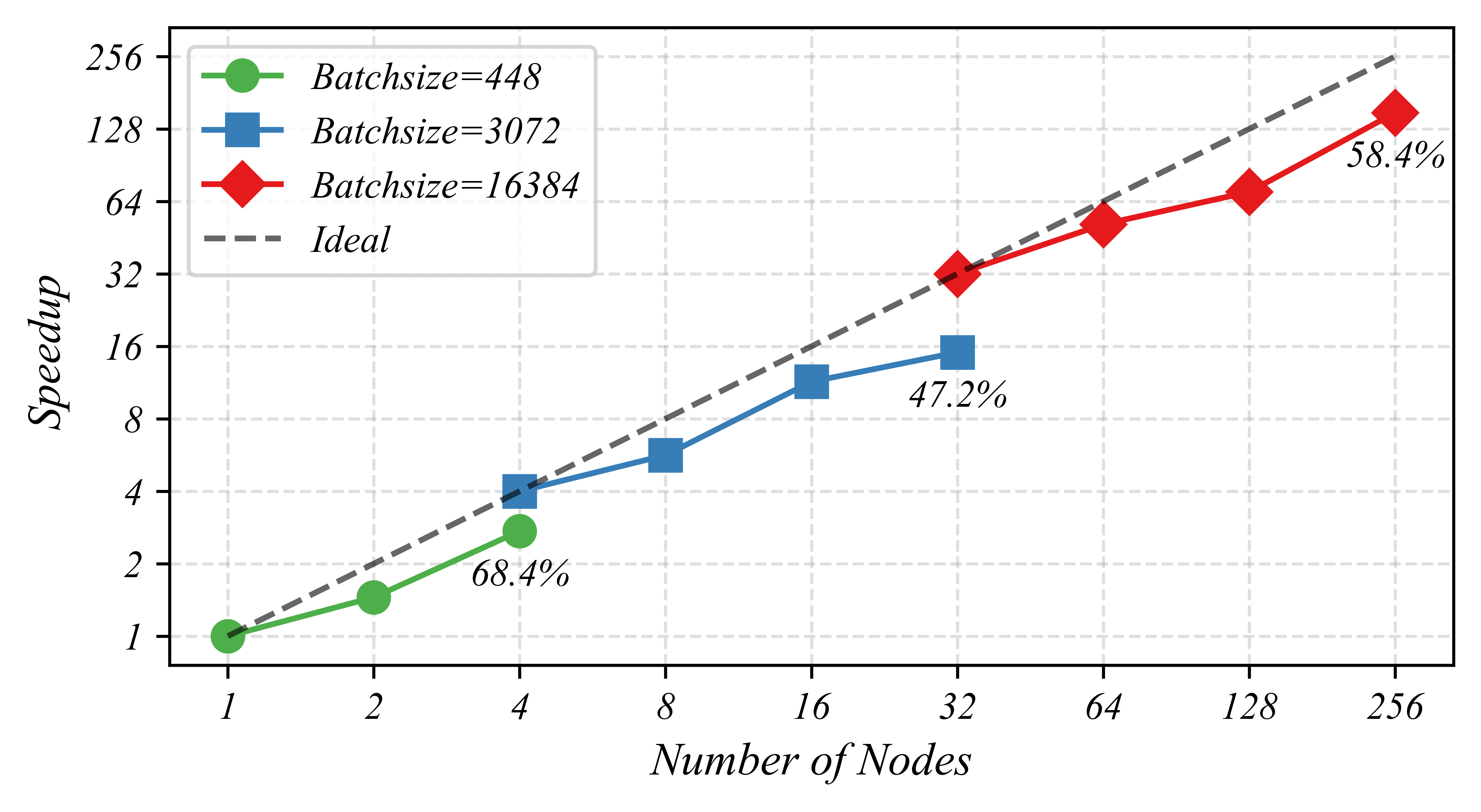}
    \captionsetup{font=small, skip=3pt}
    \caption{Strong scaling results for DiT-XL/2 training under varying batch sizes}
    \label{fig:strong_scale}
    \vspace{-0.5cm}
\end{figure}

For strong scaling, we fix the global batch size and vary the number of nodes. Guided by memory-capacity constraints, we select three global batch sizes—448, 3,072, and 16,384—and evaluate strong scaling separately for each configuration (up to 256 nodes).
As shown in Figure~\ref{fig:strong_scale}, efficiency decreases as node count increases. This is primarily because strong scaling yields a lower compute-to-communication ratio. In addition, as more nodes are added, the per-node batch size shrinks, reducing GEMM kernel efficiency.
While our optimizations effectively hide communication in the weak-scaling regime, under strong scaling the kernel-efficiency drop becomes the dominant factor.

\vspace{-0.2cm}

\section{Conclusion}

This paper presents DiT-HC, an efficient and scalable training framework for Diffusion Transformers (DiT) on HPC-oriented CPU clusters. Through systematic optimization of operator implementations, parallelization strategies, and memory scheduling, DiT-HC is able to effectively leverage new hardware features such as matrix acceleration units and hierarchical memory architectures with on-package high-bandwidth memory.
Experimental results show that DiT-HC delivers substantial improvements in end-to-end training performance and scalability. The framework achieves up to $8.2\times$ speedup over the native software stack, $87.7\times$ over OpenBLAS, and sustains more than 1 PFLOPS FP32 performance with 90.6\% weak-scaling efficiency on 256 nodes. Its effectiveness is further confirmed on real-world remote sensing datasets, where it consistently achieves the best performance across tested configurations.

Beyond these results, DiT-HC demonstrates a viable path toward narrowing the gap between GPU-centric AI training and CPU-based supercomputing. As scientific computing increasingly integrates data-driven models with traditional simulations, DiT-HC offers a promising foundation for unifying AI and HPC workflows, with potential applications in data assimilation, climate modeling, and Earth system science. Future work will extend the framework to broader classes of generative models and explore deeper integration with domain-specific scientific applications.

\vspace{-0.2cm}

\bibliography{sample-base}

\end{document}